\documentclass[english,superscriptaddress,amsmath,amssymb,notitlepage,aps,twocolumn,prl,10pt]{revtex4-2}
\usepackage[T1]{fontenc}
\usepackage{textcomp}
\usepackage[utf8]{inputenc}
\usepackage{amsbsy}
\usepackage{amssymb}
\usepackage{graphicx}

\makeatletter

\usepackage[colorlinks,linkcolor=blue, urlcolor=blue, anchorcolor=blue, citecolor=blue]{hyperref}
\usepackage{dcolumn}
\usepackage{bm}

\makeatletter
\newcommand*{\balancecolsandclearpage}{%
  \close@column@grid
  \cleardoublepage
  \twocolumngrid
}

\makeatother

\usepackage{babel}
\begin{document}
\title{Ultrafast Moiré Exciton Dynamics Revealed by Two-Dimensional Coherent
Spectroscopy}
\author{Deran Zhang}
\thanks{These authors contributed equally to this Letter.}
\affiliation{Graduate School of China Academy of Engineering Physics, No. 10 Xibeiwang
East Road, Haidian District, Beijing, 100193, China}
\author{Zhexue Chen}
\thanks{These authors contributed equally to this Letter.}
\affiliation{Suzhou Institute for Advanced Research, University of Science and
Technology of China, Suzhou 215123, China}
\affiliation{Department of Physics, University of Science and Technology of China,
Hefei, Anhui 230026, China}
\author{Dongxue Chen}
\email{dxch@ustc.edu.cn}

\affiliation{Suzhou Institute for Advanced Research, University of Science and
Technology of China, Suzhou 215123, China}
\affiliation{Department of Physics, University of Science and Technology of China,
Hefei, Anhui 230026, China}
\author{Tengfei Yan}
\email{yantf@shu.edu.cn}

\affiliation{Shanghai Engineering Research Center for Integrated Circuits and Advanced
Display Materials, School of Microelectronics, Shanghai University,
Shanghai 201800, China}
\author{Jianhua Jiang}
\email{jhjiang3@ustc.edu.cn}

\affiliation{Department of Modern Physics, School of Physical Sciences, University
of Science and Technology of China, Hefei 230026, China}
\affiliation{School of Physical Science and Technology \& Collaborative Innovation
Center of Suzhou Nano Science and Technology, Soochow University,
Suzhou 215006, China}
\affiliation{School of Biomedical Engineering, Division of Life Sciences and Medicine,
University of Science and Technology of China, Hefei 230026, China}
\affiliation{Suzhou Institute for Advanced Research, University of Science and
Technology of China, Suzhou 215123, China}
\author{Hui Dong}
\email{hdong@gscaep.ac.cn}

\affiliation{Graduate School of China Academy of Engineering Physics, No. 10 Xibeiwang
East Road, Haidian District, Beijing, 100193, China}
\date{\today}
\begin{abstract}
Moiré superlattices of van der Waals heterostructures confine excitons
into discrete energy levels in finite-sized samples, yet their internal
relaxation pathways remain obscured by inhomogeneous broadening. Here,
using two-dimensional coherent spectroscopy (2DCS), we directly resolve
seven discrete energy levels of the moiré-confined intralayer $\mathrm{WS}_{2}$
A exciton in an hBN-encapsulated $\mathrm{WS}_{2}/\mathrm{WSe}_{2}$
heterostructure, corresponding to a confinement depth of 57.02 meV.
By tracking the population of these levels over picosecond delays,
we uncover a two-stage thermalization process: an effective exciton
temperature drops from 171.8 to 99.3 K within 1.2 ps through rapid
intra-manifold population redistribution, followed by a slower recovery
toward quasi-equilibrium mediated by exciton--environment interactions.
Kullback--Leibler divergence analysis confirms the progressive transition
from a non-equilibrium to a Boltzmann-distributed population. Our
results establish state-resolved access to moiré exciton dynamics
and provide a foundation for exploring nonequilibrium many-body phenomena
in moiré quantum materials.
\end{abstract}
\maketitle
\textit{Introduction}--- Moiré superlattices in van der Waals heterostructures
provide a powerful means of engineering exciton energy landscapes
through long-period spatial modulation \citep{regan2022emerging}.
A prominent realization arises in stacked transition-metal dichalcogenide
(TMD) layers with a small twist angle or lattice mismatch, where a
small twist angle or large lattice mismatch generates a spatially
periodic potential that localizes excitons into discrete confined
states \citep{wu2017topological,yu2017moire,wu2018theory,seyler2019signatures,tran2019evidence}.
Such a moiré periodic potential offers access to quantum confinement
\citep{baek2020highly,brotons2020spin,andersen2021excitons}, exciton--exciton
interactions \citep{li2020dipolar,yu2021luminescence,zhang2021van},
and nonequilibrium dynamics \citep{yuan2020twist,kim2023dynamics,kai2025distinct}
in engineered two-dimensional materials. The staggered-gap (type-II)
band \citep{kang2013band,yu2018ultrafast} alignment further gives
rise to long-lived interlayer excitons with electrons and holes spatially
separated across the two layers, enabling strong dipolar interactions
and collective phases such as exciton condensation and superfluidity
\citep{fogler2014high,wu2015theory,wang2019evidence}. Moiré modulation
also reconstructs bright intralayer excitons into confined states
that retain large oscillator strengths, combining strong light--matter
coupling with localized and enhanced interactions \citep{wu2017topological,jin2019observation,regan2022emerging,naik2022intralayer}.

A characteristic optical signature of moiré confinement is the emergence
of multiple discrete exciton levels \citep{jin2019identification,jin2019observation,tran2019evidence,naik2022intralayer,chen2022tuning,fang2023localization,zheng2023localization}.
In near-aligned TMD heterobilayers, a single intralayer A-exciton
level has been observed to split into three prominent moiré exciton
levels \citep{jin2019observation,naik2022intralayer,chen2022tuning,fang2023localization},
yet the ladder of quantized levels---analogous to the interlayer
exciton manifold---remains unresolved \citep{tran2019evidence,jin2019identification,fang2023localization,zheng2023localization}.
Resolving this fine structure is essential both for identifying distinct
exciton species relevant to correlated exciton states \citep{xiong2023correlated,lian2024valley},
and, critically, for tracking how exciton populations redistribute
and thermalize on ultrafast timescales \citep{meneghini2024excitonic,Kim2026PRL}.
Such dynamical information has remained inaccessible because the inhomogeneous
broadening, comparable to the intrinsic homogeneous linewidth \citep{martin2020encapsulation,regan2022emerging},
smears the discrete spectral features together in conventional spectroscopy.
A measurement capable of simultaneously resolving the quantized exciton
levels and their picosecond evolution is therefore urgently demanded.

\begin{figure}[tbp]
\centering{}\includegraphics[width=8cm]{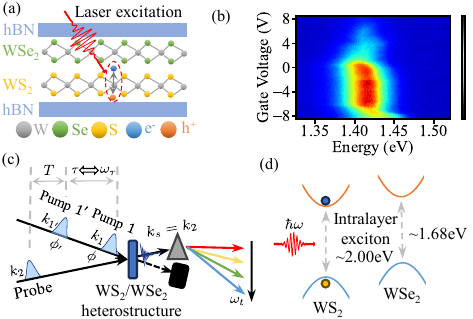}\caption{Experimental scheme and optical response of the $\mathrm{WS}_{2}/\mathrm{WSe}_{2}$
heterostructure. (a) Schematic of the hBN-encapsulated $\mathrm{WS}_{2}/\mathrm{WSe}_{2}$
heterostructure. (b) Photoluminescence spectrum of the sample measured
at $4.6\ \mathrm{K}$ under 633 nm excitation with a power of 100
$\mathrm{\mu W}$. The gate-dependent PL map shows an emission feature
near 875 nm (approximately 1416.94 meV). (c) Schematic of the two-dimensional
coherent spectroscopy measurement, with the excitation frequency $\omega_{\tau}$
obtained by Fourier transformation along the coherence time $\tau$
and the detection frequency $\omega_{t}$ recorded directly by the
spectrometer. (d) Schematic band alignment of the $\mathrm{WS}_{2}/\mathrm{WSe}_{2}$
heterostructure, illustrating the intralayer $\mathrm{WS}_{2}$ A
exciton addressed by the excitation pulses. All three pulses directly
excite the sample, with their spectra centered at 640 nm (1937.26
meV) and spanning an effective excitation bandwidth of 605-658 nm
(1884.27-2049.33 meV), covering the moiré-modulated $\mathrm{WS}_{2}$
A-exciton level. \label{fig:Num1}}
\end{figure}

In this Letter, we address this challenge using two-dimensional coherent
spectroscopy (2DCS), a purpose-built technique that disentangles inhomogeneous
broadening from the homogeneous linewidth and resolves the excitation
and emission frequencies separately \citep{mukamel1995principles,Mukamel2000ARPC,jonas2003two,tian2003femtosecond,Brixner2005,cho2009two,hamm2011concepts}.
We focus on the intralayer exciton ($\sim2.0\ \mathrm{eV}$) around
the $\mathrm{WS}_{2}$ A-exciton in an hBN-encapsulated $\mathrm{WS}_{2}/\mathrm{WSe}_{2}$
heterostructure at 81 K and resolve seven discrete levels modeled
by a moiré potential with a depth of 57.02 meV. By tracking the population
of these levels as a function of delay, we quantify the thermalization
dynamics through an effective temperature and reveal a rapid initial
decrease followed by a slower recovery on the picosecond timescale.
By simultaneously resolving the quantized exciton spectrum and its
ultrafast population redistribution, our work directly accesses the
state-resolved nonequilibrium physics of moiré excitons and provides
a route to investigate closely spaced excited states together with
their ultrafast evolution.

\textit{Moiré-modulated exciton states} --- Fig.~\ref{fig:Num1}(a)
shows the hBN-encapsulated $\mathrm{WS}_{2}/\mathrm{WSe}_{2}$ heterostructure,
with a small misalignment angle $\theta\approx1.6{}^{\circ}$. The
gate dependent photoluminescence (PL) spectrum measured at $4.6\ \mathrm{K}$
under $633\ \mathrm{nm}$ excitation with a power of 100 $\mathrm{\mu W}$
is shown in Fig.~\ref{fig:Num1}(b), providing a complementary characterization
of the heterostructure. Fig.~\ref{fig:Num1}(c) shows the experimental
measurement sequence with three phase-coherent laser pulses during
2DCS measurement. The coherence time $\tau$ is scanned and Fourier-transformed
to yield the excitation frequency $\omega_{\tau}$, typically known
as the absorption axis \citep{jonas2003two,tian2003femtosecond}.
The detection frequency $\omega_{t}$ is directly registered by a
spectrometer, serving as the emission axis. For a fixed population
delay $T$, the signal is presented with $\omega_{\tau}$ as the horizontal
axis and $\omega_{t}$ as the vertical axis. And the dynamics of the
excited states is tracked by changing $T$.

To detect the intralayer exciton, we center the excitation pulses
at 640 nm (1937.26 meV), with an effective spectral range of 605-658
nm (1884.27-2049.33 meV), directly addressing the $\mathrm{WS}_{2}$
A-excitons, as illustrated in Fig.~\ref{fig:Num1}(d). The measurements
were performed using a Ti:sapphire pulsed laser with a repetition
rate of 1 kHz and a pulse duration of approximately 35 fs. The spectroscopic
measurements are performed at 81 K, using a liquid-nitrogen-cooled
cryostat.

\begin{figure*}[tp]
\centering{}\includegraphics[width=17cm]{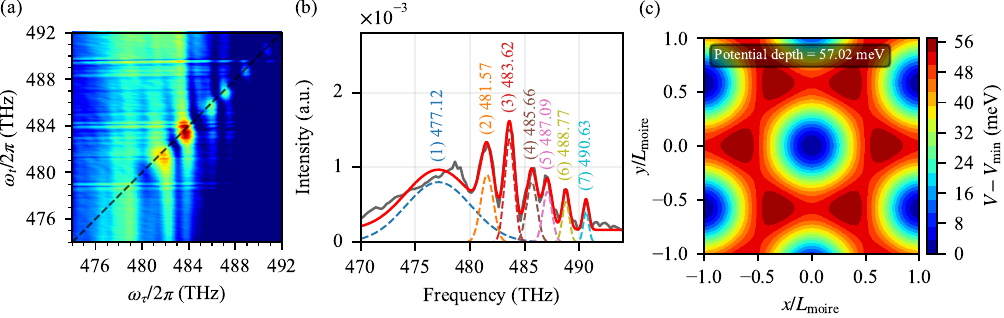}\caption{Fine energy structures of the moiré-modulated intralayer excitons
around the $\mathrm{WS}_{2}$ A-exciton. (a) Two-dimensional coherent
spectrum of the $\mathrm{WS}_{2}/\mathrm{WSe}_{2}$ heterostructure
at the population delay of $\ensuremath{T=3.2}\ \mathrm{ps}$. (b)
Diagonal cut (solid gray line) along the black-dashed line in (a).
The diagonal profile is fitted with a sum of seven Gaussian components
(dashed curves), with the total fit shown in red, yielding the central
frequencies of the seven levels at (1) $477.118\ \mathrm{THz}$ ($1973.21\ \mathrm{meV}$),
(2) $481.569\ \mathrm{THz}$ ($1991.62\ \mathrm{meV}$), (3) $483.619\ \mathrm{THz}$
($2000.10\ \mathrm{meV}$), (4) $485.661\ \mathrm{THz}$ ($2008.54\ \mathrm{meV}$),
(5) $487.088\ \mathrm{THz}$ ($2014.44\ \mathrm{meV}$), (6) $488.773\ \mathrm{THz}$
($2021.41\ \mathrm{meV}$), and (7) $490.628\ \mathrm{THz}$ ($2029.08\ \mathrm{meV}$).
(c) Real-space moiré potential obtained from the leading-harmonic
model fit. The potential is obtained from the spatially periodic potential
model \citep{wu2017topological,yu2017moire}, and shown in real-space
coordinates $x/L_{\mathrm{moire}}$ and $y/L_{\mathrm{moire}}$, where
$L_{\mathrm{moire}}$ is the moiré superlattice period, estimated
to be $\sim6.3$ nm for a twist angle of $1.62^{\circ}$. The resulting
modulation produces a potential depth of $57.02\ \mathrm{meV}$ for
the intralayer $\mathrm{WS}_{2}$ exciton. \label{fig:Num2}}
\end{figure*}

Fig.~\ref{fig:Num2}(a) shows one of the two-dimensional coherent
spectra with population delay time $T=3.2\ \mathrm{ps}$. Seven distinct
levels are resolved along the excitation--detection diagonal (black-dashed
line). Such fine structure for the intralayer exciton was typically
unresolved, in contrast to the reported ones for interlayer excitons
\citep{tran2019evidence,jin2019identification,fang2023localization,zheng2023localization}.
By identifying the positions of the diagonal peaks, we obtain seven
levels at $1973.21$, $1991.62$, $2000.10$, $2008.54$, $2014.44$,
$2021.41$, and $2029.08$ meV, as shown in Fig.~\ref{fig:Num2}(b).
The corresponding six successive energy spacings, calculated from
the fitted peak positions, are $18.41,\;8.48,\;8.44,\;5.90,\;6.97,$
and $7.67$ meV, respectively. The measured ladder exhibits nonuniform
level spacings, indicating substantial anharmonicity of the confinement
potential, consistent with the spatially periodic cosine-type moiré
potential.

Here, we model quantitatively the spatially periodic moiré potential
using its leading Fourier component \citep{wu2017topological,yu2017moire,tong2019magnetic},
\begin{equation}
V(\boldsymbol{R})=2V_{1}\sum^{3}_{j=1}\mathrm{cos}(\boldsymbol{G}_{j}\cdot\boldsymbol{R}),\label{eq:1}
\end{equation}
where \textbf{$\boldsymbol{R}$} is the center-of-mass vector. The
three symmetry-related reciprocal-lattice vectors $\boldsymbol{G}_{j}$
have the same magnitude $G$, and orientations $0^{\circ},120^{\circ},$
and $240^{\circ}$, while $V_{1}$ determines the amplitude of the
periodic modulation. The exciton eigenstates are obtained from
\begin{equation}
\left[-\frac{\hbar^{2}}{2m^{*}}\nabla^{2}+V(\boldsymbol{R})\right]\psi_{n}(\boldsymbol{R})=E_{n}\psi_{n}(\boldsymbol{R}),
\end{equation}
where $m^{*}$ represents the effective translational parameter entering
the reduced exciton Hamiltonian. Accordingly, the kinetic contribution
is parameterized by a phenomenological effective kinetic-energy scale
$E_{\mathrm{kin}}=\hbar^{2}G^{2}/(2m^{*})$, which is treated as an
independent fitting parameter of the model. Fitting the experimentally
extracted energy levels yields $E_{\mathrm{kin}}=4.93\ \mathrm{meV}$
and $V_{1}=-6.34\ \mathrm{meV}$.

The fitted moiré potential is shown in Fig.~\ref{fig:Num2}(c). The
measured energy ladder is quantitatively reproduced by the spatially
periodic moiré potential model, yielding a potential depth of 57.02
meV for the $\mathrm{WS}_{2}$ intralayer exciton. These results provide
a quantitative description of the moiré-induced confinement underlying
the observed discrete exciton levels. The extracted potential depth
is comparable in scale to the moiré potential depths reported for
excitons in $\mathrm{WSe}_{2}/\mathrm{WSe}_{2}$ homobilayers \citep{wu2022evidence}.
A similar energy scale has also emerged from first-principles calculations
of $\mathrm{WS}_{2}/\mathrm{WSe}_{2}$ heterobilayers, which predict
moiré potentials on the order of $10^{2}$ meV for interlayer excitons,
with the precise magnitude depending strongly on stacking configuration
and lattice reconstruction \citep{yuan2020twist,rossi2024anomalous}.

\textit{State-Resolved population dynamics} --- The distinct fine
energy-level structure allows us to track the dynamics for the corresponding
excitons with the variation of the population delay time $T$. We
measure the 2D spectra for 15 sets of population delay time $T$,
ranging from $0.4\ \mathrm{ps}$ to $5.0\ \mathrm{ps}$. These spectra
are shown as Fig. \ref{fig:EM1} in the End Matter.

\begin{figure}[tbp]
\centering
\centering{}\includegraphics{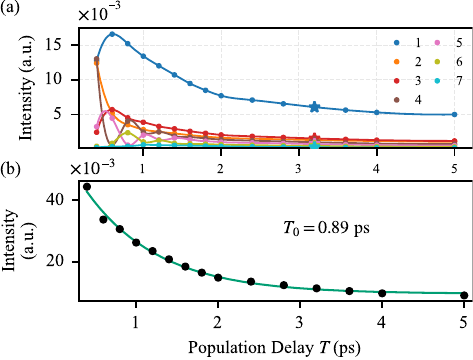}\caption{Population dynamics of the moiré-confined intralayer exciton manifold.
(a) Integrated intensities of the resolved diagonal exciton levels
as a function of the population delay time $T$. The star symbols
mark the integrated intensities at $T=3.2\ \mathrm{ps}$, used in
the analysis in Fig. \ref{fig:Num4}(a). (b) Summed intensity of the
seven resolved intralayer exciton levels as a function of population
delay. The solid curve is a fit to $I(T)=A\exp[-T/T_{0}]+C$, yielding
$T_{0}=0.89\ \mathrm{ps}$. \label{fig:Num3}}
\end{figure}

To access the dynamics of the 7 levels, we extract the peak intensities
$I_{n}(T)$, given by the integrated areas of the corresponding Gaussian
components, at each population delay time $T$ and plot their temporal
evolution as functions of $T$ in Fig.~\ref{fig:Num3}(a). The non-monotonic
dynamics of the seven exciton levels exhibit distinct temporal evolutions,
indicating a time-dependent redistribution of spectral weight among
the discrete moiré exciton states. Concurrently, the summed intensity
of the seven resolved intralayer levels decreases over the same delay
range, as shown in Fig.~\ref{fig:Num3}(b). Its decay is well described
by a single-exponential component, $I(T)=A\mathrm{exp}[-T/T_{0}]+C,$
yielding a characteristic time $T_{0}=0.89$ ps.

\begin{figure}[tbp]
\centering{}\includegraphics[width=8cm]{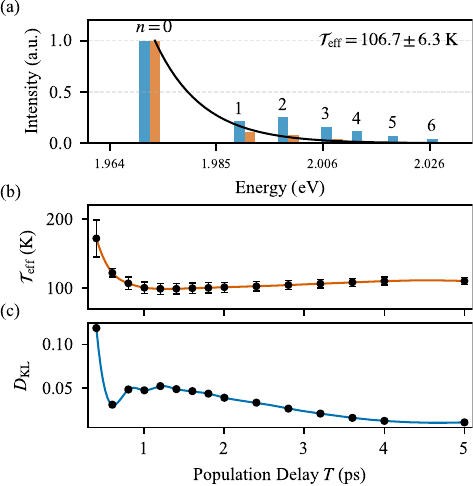}\caption{Thermalization of the moiré-confined intralayer exciton manifold.
(a) Integrated intensities (blue bars) and degeneracy-corrected intensities
(orange bars) of the diagonal exciton states at $T=3.2\ \mathrm{ps}$,
corresponding to the star-marked delay in Fig.~\ref{fig:Num3}(a). The solid curve
is a Boltzmann fit to the degeneracy-corrected intensities, yielding
$\mathcal{T}_{\mathrm{eff}}=106.7\pm6.3\ \mathrm{K}$. (b) Effective
temperature $\mathcal{T}_{\mathrm{eff}}(T)$ extracted at each population
delay $T$. Error bars denote the fitting uncertainties. (c) Kullback--Leibler
divergence $D_{\mathrm{KL}}(P_{n,\mathrm{norm}}\Vert P^{\mathrm{eq}}_{n,\mathrm{norm}})$
between the experimentally determined population distribution $P_{n}$
and the corresponding Boltzmann distribution $P^{\mathrm{eq}}_{n}$.
Smaller $D_{\mathrm{KL}}$ indicates closer agreement with a Boltzmann
distribution. The solid curves in (b) and (c) are cubic-spline interpolations.
\label{fig:Num4}}
\end{figure}

Taking into account their degeneracies $g_{n}=n+1$ for the state
$n$, we get the population $P_{n}(T)$ of the individual states as
\begin{equation}
P_{n}(T)\propto\frac{I_{n}(T)}{g_{n}}.
\end{equation}
Fig.~\ref{fig:Num4}(a) displays the degeneracy-corrected population
distribution at $T=3.2\ \mathrm{ps}$ with orange bars, with the blue
bars representing the total intensities of the seven levels.

To quantify the distribution, we fit the data with a Boltzmann distribution
with an effective temperature $\mathcal{T}_{\mathrm{eff}}(T)$ as
\begin{equation}
P^{\mathrm{eq}}_{n}(T)\propto\mathrm{exp}\left[-\frac{E_{n}}{k_{B}\mathcal{T}_{\mathrm{eff}}(T)}\right].\label{eq:4}
\end{equation}
The black line in Fig.~\ref{fig:Num4}(a) shows this fitting curve.
Fitting all the data at different delay times $T$, we obtain the
effective temperature as a function of $T$, illustrated in Fig.~\ref{fig:Num4}(b).
Within approximately the initial 1.2 ps, the effective temperature
decreases from 171.8 K to a minimum of 99.3 K, corresponding to a
reduction of 72.5 K ($\sim42\%$) from its initial value. At later
times, $\mathcal{T}_{\mathrm{eff}}$ increases to 110.4 K at 5.0 ps.

We next assess the conformity of the measured population distribution
to the corresponding Boltzmann distribution. Fig.~\ref{fig:Num4}(c)
shows the Kullback-Leibler divergence $D_{\mathrm{KL}}(P_{n,\mathrm{norm}}\Vert P^{\mathrm{eq}}_{n,\mathrm{norm}})$,
which provides a nonnegative measure of deviations of the normalized
experimental population distribution from the corresponding equilibrium
distribution. Detailed definitions are given in the END MATTER. Such
divergence $D_{\mathrm{KL}}$ shows a weak nonmonotonic variation
during the initial $1.2$ ps, and a subsequent decrease after 1.2
ps. Such dynamics of the Kullback-{}-Leibler divergence illustrates
the transition from a highly non-equilibrium distribution to a quasi-equilibrium
Boltzmann distribution.

The population dynamics shows two dynamical regimes, a short-time
nonmonotonic reduction and a long-time decay. Firstly, the short-time
decay timescales (0.89 ps) of the summed intralayer-exciton intensity
is comparable to the timescale of the rapid variation of $\mathcal{T}_{\mathrm{eff}}$
(1.2 ps). These comparable timescales identify an early dynamical
regime in which substantial spectral-weight loss from the resolved
intralayer exciton states occurs simultaneously with strong redistribution
among these states. Several microscopic channels may contribute on
this timescale. Interlayer charge transfer \citep{yu2018ultrafast,wei2025spatiotemporal}
and spin-forbidden dark exciton states \citep{plechinger2016trion,selig2016excitonic}
or momentum-indirect exciton states \citep{wallauer2021momentum}
may also be related to the depletion of the optically resolved intralayer
population, while many-body exciton--exciton interactions and optical-phonon
emission can facilitate rapid redistribution and energy relaxation
within the exciton population.

Secondly, during the later stage, the exciton dynamics are characterized
by a continued loss of excitation from the resolved moiré-state manifold
together with a progressive evolution toward a near-equilibrium population
distribution. These slower dynamics are consistent with relaxation
mediated by exciton--environment interactions, including coupling
to the lattice and other environmental degrees of freedom \citep{ovesen2019interlayer,bange2023ultrafast,Kim2026PRL}.

\textit{Conclusion} --- Taking advantage of the state-resolving and
time-domain capabilities of two-dimensional coherent spectroscopy,
we have resolved the seven-level quantized structure of the moiré-confined
$\mathrm{WS}_{2}$ A-exciton---unresolved in prior work because inhomogeneous
broadening is comparable to the homogeneous linewidth---and tracked
its picosecond population dynamics in a $\mathrm{WS}_{2}/\mathrm{WSe}_{2}$
heterostructure. A spatially periodic moiré potential model yields
a confinement depth of $57.02$ meV, comparable to that of interlayer
excitons \citep{yuan2020twist,rossi2024anomalous}. The population
dynamics reveal a two-stage thermalization: an initial rapid drop
of the effective exciton temperature from 171.8 to 99.3 K within $1.2$
ps, driven by intra-manifold redistribution, followed by a slower
recovery mediated by exciton--environment coupling. Kullback--Leibler
divergence analysis confirms the progressive transition from a non-equilibrium
to a quasi-equilibrium Boltzmann distribution. These results establish
2DCS as a direct window into the nonequilibrium internal dynamics
of moiré excitons and lay the groundwork for exploring many-body physics
in moiré quantum materials.
\begin{acknowledgments}
This work is supported by the Innovation Program for Quantum Science and Technology (Grant No. 2023ZD0300700), the National Natural Science Foundation of China (Grant Nos. U2230203, U2330401, 12088101, 12574348, 52503407), and Natural Science Foundation of Anhui Province (Grant Numbers 2408085MF176). We would like to thank Dr. Wang Yao for the helpful discussion.
\end{acknowledgments}

\bibliographystyle{apsrev4-2}
\bibliography{ref}

\appendix
\clearpage

\section*{End Matter}

\begin{figure*}[tp]
\centering{}\includegraphics[width=17cm]{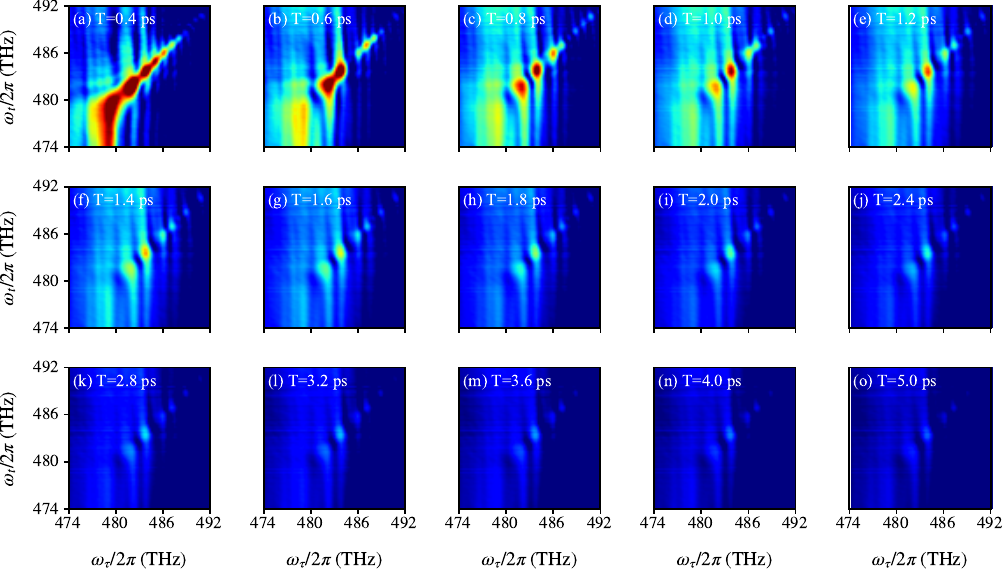}\caption{Population-delay evolution of the two-dimensional coherent spectra.
(a)--(o) Rephasing 2D spectra measured at population delays of $T=0.4,\ 0.6,\ 0.8,\ 1.0,\ 1.2,\ 1.4,\ 1.6,\ 1.8,\ 2.0,\ 2.4,\ 2.8,\ 3.2,\ 3.6,\ 4.0$,
and $5.0$ ps, respectively. The horizontal and vertical axes denote
the excitation and detection frequencies, respectively. All panels
use the same color scale. The diagonal levels reveal the temporal
evolution of the moiré-modulated WS\textsubscript{2} A-exciton spectrum
and are used for the state-resolved dynamical analysis. \label{fig:EM1}}
\end{figure*}

\textit{Sample preparation and twist-angle determination}--- The
preparation of the moiré superlattice employed a stacking approach
based on van der Waals dry transfer technology. First, bulk $\mathrm{WS}_{2}$,
$\mathrm{WSe}_{2}$, hexagonal boron nitride (hBN), and graphite were
individually exfoliated onto PDMS using mechanical exfoliation. The
various materials were then transferred separately onto $\mathrm{SiO_{2}/Si}$
substrates, followed by vacuum annealing at 300 °C for 10 h to remove
residual adhesive. Monolayer $\mathrm{WS}_{2}$ and $\mathrm{WSe}_{2}$
were confirmed using optical microscopy combined with Raman spectroscopy.
After all materials were prepared, layer-by-layer stacking was performed
using a transparent polycarbonate/polydimethylsiloxane (PC/PDMS) stamp.
During the stacking process, the rotation angle was precisely controlled
using a two-dimensional (2D) material transfer platform to obtain
$\mathrm{WS}_{2}/\mathrm{WSe}_{2}$ moiré superlattice with a specific
angle. Few-layer graphene serves as the gate, while few-layer hBN
encapsulates and protects the moiré superlattice. A Ti/Au bottom electrode
was fabricated using laser direct writing, and the stacked materials
were subsequently released onto the bottom electrode on the 2D transfer
platform.

The relative crystallographic orientation of the $\mathrm{WS}_{2}$
and $\mathrm{WSe}_{2}$ monolayers was independently determined by
polarization-resolved second-harmonic generation (SHG), yielding a
twist angle of approximately $1.62^{\circ}$. See the Supplementary
Materials for details. The moiré superlattice period was estimated
by accounting for both the lattice mismatch and the relative twist
angle,
\begin{equation}
L_{\mathrm{moire}}=\frac{a_{\mathrm{WS_{2}}}a_{\mathrm{WSe_{2}}}}{\sqrt{a^{2}_{\mathrm{WS_{2}}}+a^{2}_{\mathrm{WSe_{2}}}-2a_{\mathrm{WS_{2}}}a_{\mathrm{WSe_{2}}}\cos\theta}},
\end{equation}
where $a_{\mathrm{WS_{2}}}=0.318\ \mathrm{nm}$ and $a_{\mathrm{WSe_{2}}}=0.332\ \mathrm{nm}$
are the lattice constants of monolayer $\mathrm{WS}_{2}$ and $\mathrm{WSe}_{2}$,
respectively, and $\theta=1.62^{\circ}$ is the measured relative
twist angle. These parameters yield $L_{\mathrm{moire}}\approx6.3\ \mathrm{nm}$.

\textit{Two-dimensional coherent spectroscopy and data processing}
--- For each population delay $T$, the coherence time $\tau$ was
scanned from 0 to 1.0 ps in 5.0 fs steps. A four-step phase-cycling
sequence with relative phases (0,0), (0,$\pi$/2), (0,$\pi$),
and (0,3$\pi$/2) was used to reconstruct the rephasing
signal presented throughout this work \citep{zhang2012phase}. This
photon-echo-type response arises from opposite phase evolution during
the coherence and detection intervals, allowing rephasing of optical
coherences dephased by static inhomogeneity \citep{jonas2003two}.
Fourier transformation along $\tau$ yields the excitation-frequency
axis $\omega_{\tau}$. Fig.~\ref{fig:EM1} shows the rephasing 2D
spectra measured from T = 0.4 to 5.0 ps in panels (a)--(o). At the
sample position, the powers of pump 1 and pump 1\textasciiacute{}
were both 24.28 $\mathrm{\mu W}$, with an estimated spot diameter
of approximately 140 $\mathrm{\mu m}$, while the probe power was
0.34 $\mathrm{\mu W}$, with a spot diameter of approximately 27 $\mathrm{\mu m}$.
Seven diagonal levels remain distinguishable throughout the measured
range, revealing the internal structure of the moiré-confined $\mathrm{WS}_{2}$
A-exciton states. The spectra show pronounced spectral-weight redistribution
within approximately 1.0 ps, followed by slower evolution over several
picoseconds, motivating the effective-temperature and population-distribution
analyses in the main text.

We also perform the measurement on two different samples. The spectra
exhibit consistent excitonic fine structure and similar dynamical
behavior. These data are not shown here but are available from the
corresponding authors upon reasonable request.

\textit{Extraction and quantitative modeling of the resolved exciton
states}--- For each population delay T, the diagonal profile was
extracted from the magnitude of the rephasing two-dimensional spectrum
and fitted by the sum of seven Gaussian components together with a
constant background,
\begin{equation}
S(f,T)=C+\sum^{7}_{n=1}A_{n}(T)\exp[-(f-f_{n})^{2}/(2\sigma^{2}_{n})],
\end{equation}
where $f_{n}$, $A_{n}$, and $\sigma_{n}$ denote the center frequency,
amplitude, and linewidth of the n-th level, respectively. The level
energies were determined from the fitted center frequencies. The $1\sigma$
uncertainties of the fitted Gaussian parameters were obtained from
the diagonal elements of the covariance matrix returned by the nonlinear
least-squares fit.

The intensity $I_{n}(T)$ of each resolved level was determined from
the integrated area of the corresponding fitted Gaussian component,
yielding $I_{n}=A_{n}\sigma_{n}\sqrt{2\pi}$. The Gaussian components
were numerically integrated over the fitting range from 450 to 510
THz. The same fitting and integration procedure was applied at each
population delay to obtain the state-resolved intensities used in
the dynamical analysis. The summed intensity shown in Fig.~\ref{fig:Num3}(b)
was calculated as $I_{\mathrm{tot}}(T)=\sum^{7}_{n=1}I_{n}(T)$.

For the quantitative determination of the moiré energy ladder, the
seven fitted center frequencies were converted into relative energies
with respect to the lowest-energy level and compared with the eigenenergies
of the spatially periodic moiré potential introduced in the main text.
The exciton center-of-mass Hamiltonian was diagonalized at the $\Gamma$
point using a converged plane-wave basis. The calculated eigenvalues
were grouped into successive shells containing 1, 2, ..., 7 states
according to the harmonic-shell assignment, and the mean eigenenergy
of each shell was used as the corresponding shell energy.

The phenomenological effective kinetic-energy scale $E_{\mathrm{kin}}$
and the leading Fourier component $V_{1}$ were optimized by a global
search followed by nonlinear least-squares fitting to the six excited-state
energies. The fitted-minus-experimental deviations for the seven levels
are 0, -6.08, -3.32, -1.51, +0.10, +2.55, and +1.24 meV, respectively,
yielding RMSE=3.12 meV. The real-space potential obtained from the
fitted Fourier components has a potential depth $V_{\max}-V_{\min}=57.02$
meV for the intralayer $\mathrm{WS}_{2}$ exciton.

\textit{Degeneracy correction and thermalization analysis}--- The
shell multiplicity assigned to each resolved exciton level follows
the two-dimensional harmonic-oscillator structure obtained near a
moiré-potential minimum. Within this harmonic approximation, the ground
state is associated with $(n_{x},n_{y})=(0,0)$, while the first excited
shell contains the degeneracy (1,0) and (0,1). More generally, states
satisfying $n_{x}+n_{y}=n$ form the nth excitation shell, giving
an effective shell multiplicity $g_{n}=n+1$. For the  periodic potential,
anharmonicity can lift the exact degeneracy within a shell, and $g_{n}$
is therefore used here as the effective shell multiplicity associated
with this harmonic-limit assignment. Accordingly, the experimentally
extracted integrated intensity $I_{n}(T)$ was corrected by the corresponding
degeneracy, and the normalized population was defined as
\begin{equation}
P_{n,\mathrm{norm}}(T)=\frac{I_{n}(T)/g_{n}}{\sum_{m}I_{m}(T)/g_{m}}.
\end{equation}
The degeneracy-corrected populations at each population delay were
fitted with a Boltzmann distribution,
\begin{equation}
P^{\mathrm{eq}}_{n,\mathrm{norm}}(T)=\frac{\exp[-(E_{n}-E_{0})/(k_{B}\mathcal{T}_{\mathrm{eff}})]}{\sum_{m}\exp[-(E_{m}-E_{0})/(k_{B}\mathcal{T}_{\mathrm{eff}})]},
\end{equation}
from which the effective temperature $\mathcal{T}_{\mathrm{eff}}$
was extracted. To quantify how closely the measured population distribution
approaches the corresponding Boltzmann distribution, we further calculated
the Kullback--Leibler divergence,
\begin{equation}
D_{\mathrm{KL}}(P_{n,\mathrm{norm}}\Vert P^{\mathrm{eq}}_{n,\mathrm{norm}})=\sum_{n}P_{n,\mathrm{norm}}\ln\frac{P_{n,\mathrm{norm}}}{P^{\mathrm{eq}}_{n,\mathrm{norm}}}.
\end{equation}
Both distributions were normalized independently at each population
delay. $D_{\mathrm{KL}}$ is nonnegative and approaches zero as the
experimental population distribution approaches the fitted Boltzmann
distribution. This quantity therefore provides a complementary measure
of thermalization that is complementary to the fitted effective temperature
$\mathcal{T}_{\mathrm{eff}}$.
\end{document}